\documentclass[11pt,twoside]{article}
\usepackage{osid}
\usepackage{amsmath}
\usepackage{float}
\usepackage{longtable}
\usepackage{epsfig}
\usepackage{amssymb}
\usepackage{amsfonts}
\usepackage{latexsym}
\usepackage{theorem}
\usepackage{times}
\usepackage{bm}
\usepackage[pdftex]{color}
\usepackage[cmtip,arrow]{xy}
\usepackage{pb-diagram,pb-xy}
\usepackage{verbatim}
\usepackage{fancyhdr}
\usepackage{tikz}
\usepackage{amscd}
\def\1#1{{\bf #1}}
\def\2#1{{\mathcal #1}}
\def\4#1{{\tt #1}}
\def\5#1{{\sf #1}}
\def\6#1{{\mathfrak #1}}
\def\7#1{{\Bbb #1}}
\def\8#1{{\rm #1}}
\def\9#1{{\mathcurl #1}}

\def\RHD{D.-M. Schlingemann, Remarks on the structure of Clifford quantum cellular automata}
\def\LHD{D.-M. Schlingemann, Remarks on the structure of Clifford quantum cellular automata}
\DeclareFontFamily{OT1}{rsfs}{}
\DeclareFontShape{OT1}{rsfs}{m}{n}{<-7> rsfs5 <7-10> rsfs7 <10-> rsfs10}{}
\DeclareMathAlphabet\mathcurl{OT1}{rsfs}{m}{n}

\title{Remarks on the structure of Clifford quantum cellular automata}
\author{Dirk-Michael Schlingemann \\
{\footnotesize\it ISI Foundation, Quantum Information Theory Group\\
Viale S. Severo 65\\
10133 Torino, Italy}}
\begin{document}
\maketitle
\begin{abstract}
We report here on the structure of reversible quantum cellular automata with the additional
restriction that these are also Clifford operations. This means that tensor products of Weyl
operators (projective representation of a finite abelian symplectic group) are mapped to multiples
of tensor products of Weyl operators. Therefore Clifford quantum cellular automata are induced
by symplectic cellular automata in phase space. We characterize these symplectic cellular automata
and find that all possible local rules must be, up to some global shift, reflection invariant with
respect to the origin. In the one dimensional case we also find that all 1D Clifford quantum
cellular automata are generated by a few elementary operations. 
\end{abstract}

\section{Introduction}
A standard modeling technique for various complex systems are cellular automata. In fact they are
ideally suited for models of diverse phenomena as coffee percolation, highway traffic and oil
extraction from porous media. Cellular automata also provide an abstract computational model
that can simulate Turing machines, and even explicit simple automata such as Conway's life game have
been shown to support universal computation \cite{Conw76}. Quantum cellular automata provide model
for analyzing quantum computational processes and quantum computational complexity. In his famous
paper \cite{Fey82}, Feynman discusses this idea in order to obtain a model for quantum computing
which can be more powerful than a classical computer. Quantum cellular automata also play a role in
view of quantum computational complexity. This aspect has been studied by Bernstein and Vazirani
\cite{BerVaz93} by using the concept of quantum Turing machines. Watrous (see e.g. \cite{Wat95})
continued this discussion by relating quantum Turing machines to quantum cellular automata. Last but
not least, there may also be interesting applications besides quantum information theory, for
instance, quantum cellular automata could serve as ultra-violet regularized quantum field theories. 

In this paper, we are concerned with  the ``quantized version'' of cellular automata based on the
concepts that are outlined in the article of Schumacher and Werner \cite{SchuWer04}. 

\section{The general concept}
In order to motivate the concept for ``quantum'' cellular automata, we briefly review here the idea
of reversible classical cellular automata from an algebraic point of view. In this context a complex
classical system consists of single cells that are labeled by a countable set $X$. For many
applications this set is given by a regular cubical lattice $X=\7Z^d$ of dimension $d$. In our later
discussion, we restrict our considerations to this case. To each cell $x\in X$ a finite set of
states $Q$ is associated, so that the classical configuration space of the total system is given by
all functions from $X$ into the set $Q$. A further important aspect is concerned with the local
action of cellular automata which means that when the automaton is applied, the updated state of a
single cell $x\in X$ only depends on the states of a finite set $N(x)\subset X$ of  ``neighboring''
cells. In the case of a regular cubic lattice $X=\7Z^d$, the neighboring scheme is usually chosen to
be translationally invariant. Here one takes some finite set $N\subset \7Z^d$ and defines the
neighbors of a cell $x$ according to $N(x)=N+x$. In mathematical terms, a reversible classical
quantum cellular automata is given a bijective map $T\mathpunct: Q^X\to Q^X$ such that for each
$x\in X$ there exists a function $T_x\mathpunct:Q^{N(x)}\to Q$ which satisfies
$T(q)(x)=T_x(q|_{N(x)})$. Here $q$ is a function in $Q^X$ and $q|_{N(x)}$ denotes the restriction
of $q$ to the neighborhood $N(x)$.

In the case of a regular lattice structure $X=\7Z^d$, translation invariance is an additional
requirement for the automaton $T$. The translation group $\7Z^d$ acts naturally on the
configuration space $Q^{\7Z^d}$ according to $(\tau_x q)(y)=q(y-x)$. Translation invariance for $T$
means that $T$ commutes with all translations $\tau_x$. As a consequence the cellular automaton $T$
is completely determined by its ``local rule'' $T_0\mathpunct:Q^N\to Q$. Namely, the local rule
$T_x$ at any cell $x$ can be calculated from the local rule $T_0$ at the origin by
$T_x(q|_{N+x})=T_0(\tau_{-x}q|_{N})$. 

In order to motivate the quantized concept, we reformulate the classical quantum cellular automata
algebraically. If $Q$ is a finite set, then the configuration space $Q^X$ is compact in
the Tychonov topology. The observable algebra of the classical system is given by the abelian
C*-algebra of continuous functions $C(Q^X)$ which is canonically isomorphic to the tensor product
$C(Q^X)=\otimes_{x\in X} C(Q)$, where $C(Q)\cong \7C^Q$ is the abelian C*-algebra of functions on
the single cell configuration space $Q$. In this picture, a reversible cellular automaton $T$
induces an automorphism  $\alpha$ on $C(Q^X)$ by the pullback $\alpha(f)(q)=f(T(q))$. 

The functions in $C(Q^X)$ with values in the interval $[0,1]$ can be regarded as classical
observables. Such an observable is localized in a subset $U\subset X$ if the corresponding function
$f$ only depends on the restriction $q|_U$ of a classical configuration $q$. With abuse of notation
we express this fact as $f(q)=f(q|_U)$. Thus if we restrict the automorphism $\alpha$ to observables
that are localized in a single cell $x$ the resulting observable is localized in the neighbor
hood of $x$. This can be verified as follows: If an observable $f$ is localized at $x$, then the
value $f(q)=f(q(x))$ only depends on $q(x)$. The application of the automorphism $\alpha$
therefore gives $\alpha(f)(q(x))=f(T(q)(x))=f(T_x(q|_{N(x)})$, where $T_x$ is the local rule at $x$.
Thus the automorphism $\alpha$ propagates the localization region of an observable only into its
neighboring cells.

The basic idea to quantize the concept of cellular automata is to replace the classical systems by
quantum systems, i.e. the abelian C*-algebra $C(Q^X)=\otimes_{x\in X}C(Q)$ is replaced be a 
non-abelian one. The system under consideration is now described by a tensor product 
\begin{equation}
\6A=\bigotimes_{x\in X} \6A(x) \; 
\end{equation}
where to each cell $x$ there is a (finite dimensional) C*-algebra $\6A(x)$ associated with. We
require here, that to each cell $x$ an isomorphic copy of a fixed C*-algebra is assigned, i.e.
$\6A(x)\cong \6A_0$. Since $X$ can be any countable set, we have to deal here with infinite tensor
products. However, this is well defined in terms of the so called inductive limit. The algebra
$\6A$ is usually called the ``quasi-local'' algebra of observables. For each finite subset $U\subset
X$ there is a natural ``local'' subalgebra $\6A(U)=\otimes_{x\in U}\6A(x)$ of the quasi local
algebra. The operators $a\in\6A(U)$ are identified with operators in $\6A(X)$ by filling the
remaining tensor positions $X\setminus U$ with the unit operator. The concept of a reversible
quantum cellular automaton is defined as follows:

\begin{definition}{Definition}
\label{def:qca}
A reversible quantum cellular automata (QCA) is a *-automorphism $\alpha$ of the quasi local algebra
$\6A$ that fulfills the ``locality condition'': For each cell $x\in X$ and for each operator
$A\in\6A(x)$ the operator $\alpha(A)\in \6A(N(x))$ is localized in the neighborhood of $x$.
\end{definition}

As Definition~\ref{def:qca} indicates, the concept of a QCA works for any type of lattice $X$
with an appropriate neighborhood scheme where the locality requirement is the essential
ingredient. In the subsequent analysis we focus on regular cubic lattices $X=\7Z^d$ only. In this
case we have a natural action of the lattice translation group $\7Z^d$ by automorphisms $\tau_x$ on
the quasilocal algebra where $\tau_x$ is determined by 
\begin{equation}
\tau_x\left(\bigotimes_{y\in X} A_y\right)=\bigotimes_{y\in X} A_{y-x}
\end{equation}
with $A_x\in\6A(x)$. We now consider those QCAs that respect the symmetry of lattice
translations. 

\begin{definition}{Definition}
\label{def:qca_invariant}
A translationally invariant reversible QCA is a reversible QCA $\alpha$ that commutes with the
lattice translation group:  $\alpha\circ \tau_x=\tau_x\circ \alpha$.
\end{definition}

In the subsequent we always refere to the translationally invariant situation. The translation
symmetry can be exploited for the structural analysis of QCAs. In fact, a translationally invariant
QCA is completely determined by its local rule at the origin. Recall that the local rule
$\alpha_0$ at $x=0$ is the restriction of the QCA $\alpha$ to the algebra $\6A(0)$. Due to the
locality condition, there is a finite subset $N\subset \7Z^d$ such that
$\alpha(\6A(0))\subset\6A(N)$. To be compatible with translation invariance, the neighborhood scheme
can be chosen such that $N(x)=N+x$ and the ``global rule'', which is just the automorphism 
$\alpha$, can be expressed in terms of the local rule $\alpha_0$ by
\begin{equation}
\alpha\left(\bigotimes_{x\in X} A_x\right)=\prod_{x\in X} \tau_{x}\alpha_0\tau_{-x}(A_x)\; .
\end{equation}
Thus a translationally invariant QCAs can be described in terms of its local rule only. In
particular, if the single cell algebras are finite dimensional, the consruction of the QCA is a
problem in finite dimensions. 

A strategy for constructing a QCA is based on finding a valid local rule. One has to choose a
*-homomorphism $\alpha_0\mathpunct:\6A(0)\to \6A(N)$ and has to check the commutator condition
$[\tau_x(\alpha_0(A)),\alpha_0(B)]=0$ for all $A,B\in\6A(0)$ and for all $x\in X$ with $N\cap
N+x\not=\emptyset$. Assuming that the single cell algebra $\6A(0)$ is finite dimensional, there
are only finitely many conditions to be tested. For a comprehensive review on this issue, we refere
here to the work of Schumacher and Werner \cite{SchuWer04}.

Although there are only finitely many conditions to check, a general systematic classification of
QCAs is a highly non-trivial and still unsolved task. But there are particular classes of QCAs for
which a complete and explicit classification is possible, as the class of Clifford (or quasifree)
quantum cellular automata which we review here in the following. The results that we are presenting
here are based on our previous article \cite{SchlVogtWer08}.

\section{Clifford quantum cellular automata}
To explain the concept of Clifford quantum cellular, we consider a regular cubic lattice $\7Z^d$.
To each cell we associate a full matrix algebra $\6A(x)=\8M_p(\7C)$ where $p$ is a prime
number. Moreover we choose a basis of Weyl operators in $\8M_p(\7C)$. These operators are
generalizations of the Pauli operators and are constructed by shift and multiplier unitaries. To be
more precise, we consider an orthonormal basis $|q\rangle$ of the Hilbert space $\7C^p$ that is
labled by elements $q$ of the finite field $\7F_p=\7Z_p$. One way to define the Weyl operators is
to determine its action on the basis $|q\rangle$ according to 
\begin{equation}
\1w(\xi)|q\rangle=\1w(\xi_+,\xi_-)|q\rangle=\varepsilon_p^{\xi_+q}|q+\xi_-\rangle
\end{equation}
where $\varepsilon_p$ is the $p$th root of unity. As a consequence, the Weyl operators fulfill the
relation
\begin{equation}
\1w(\xi+\eta)=\varepsilon_p^{\xi_-\eta_+}\1w(\xi)\1w(\eta)
\end{equation}
which shows that the Weyl operators form a unitary projective representation of the additive group
$\7F_p$. 

For the special case $p=2$, which corresponds to qubits, the corresponding Weyl operators are
related to the Pauli matrices $X,Y,Z$ by $X=\1w(0,1)$, $Z=\1w(1,0)$ and $Y=-\8i\1w(1,1)$. 

We are now concerned with the quasi local algebra $\6A$ which is given by the infinite tensor
product of single cell algebras $\6A(x)=\8M_p(\7C)$ over the regular lattice $\7Z^d$. The
Weyl operators for the infinite system are given by tensor products  
\begin{equation}
\1w(\xi):=\bigotimes_{x\in\7Z}\1w(\xi(x))
\end{equation}
where the ``phase space'' vector $\xi$ is a function from the lattice $\7Z^d$ to the vector space
$\7F_p^2$ with finite support. Note that the finite support condition guarantees that only
finitely many tensor factors are different from the identity which implies that $\1w(\xi)$ is a
well defined unitary operator that belongs to quasilocal algebra. Moreover, the complex linear hull
of the Weyl operators is norm dense subalgebra of $\6A$. I this sense, the Weyl operators form a
``basis'' for the quasi local algebra. 

Obviously, the Weyl operators of the infinite system fulfill
the relation 
\begin{equation}
\1w(\xi+\eta)=\epsilon_p^{\beta(\xi,\eta)}\1w(\xi)\1w(\eta) \; \mbox{ with } \;
\beta(\xi,\eta)=\sum_{x\in\7Z}\xi_+(x)\eta_-(x) \; .
\end{equation}
which implies the commutation relation
\begin{equation}
\1w(\eta)\1w(\xi)=\epsilon_p^{\sigma(\xi,\eta)}\1w(\xi)\1w(\eta) \; \mbox{ with } \;
\sigma(\xi,\eta)=\beta(\xi,\eta)-\beta(\eta,\xi)\; .
\end{equation}
Note that the symplectic form $\sigma$ for the infinite system is well defined since it is evaluated
only for function with finite support. This relations justifies to interprete the functions $\xi$
as vectors in a discrete phase space --- denoted by $\Xi_{p,d}$ in the following --- with symplectic
form $\sigma$. We are now prepared to give a precise definition of Clifford quantum cellular
automata.

\begin{definition}{Definition}
A Clifford quantum cellular automata (CQCA) $\alpha$ is a translationally invariant reversible QCA
which maps Weyl operators to multiples of Weyl operators. Thus there exists a function
$\1S\mathpunct:\Xi_{p,d}\to\Xi_{p,d}$ as well as a phase-valued function
$\varphi\mathpunct:\Xi_{p,d}\mapsto \8U(1)=\{z\in\7C||z|=1\}$
such that
\begin{equation}
\label{eq:CLIFF}
\alpha(\1w(\xi))=\varphi(\xi)\1w(\1S\xi)
\end{equation}
holds for all phase space vectors $\xi$.
\end{definition}

A simple CQCA is given by lattice translations. The lattice translations act on the phase space
vectors $\xi$ by just translating the function $(\tau_y\xi)(x)=\xi(x-y)$. With abuse of notation we
use the same symbol for the action on phase space as for the action on the quasilocal algebra. By
construction the covariance relation
\begin{equation}
\tau_x(\1w(\xi))=\1w(\tau_x\xi)
\end{equation}
follows immediately. Hence, the lattice translations $\tau_x$ are CQCAs in the sense of
the definition given above.

As we will see, the translation invariance together with the condition to map Weyl operators to
multiple of Weyl operators is sufficient to determine a CQCA. This means, that the locality is
a consequence of these conditions. To deal with the translation symmetry in an appropriate way, we
identify the phase space $\Xi{p,d}=D_{p,d}^2$ as a two dimensional module over the ring $D_{p,d}$ of
functions from the lattice $\7Z^d$ into the finite field $\7F_p$ having finite support. The
multiplication in the ring is the convolution of functions which is given by 
\begin{equation}
f\star g=\sum_x f(x)\tau_x g \, .
\end{equation}
We are now prepared to state the first structure theorem on CQCAs. We refere here the reader to our
article \cite{SchlVogtWer08} for a complete discussion of the proof in which uses techniques from
the theory of projective representations of symplectic abelian groups (see e.g. \cite{Zmud72}) as
well as results from the theory of covariant completely positive maps \cite{Scu79,Hol04}.

\begin{theorem}{Theorem}
For each CQCA $\alpha$, there exists a two-by-two matrix
$\1s\in\8M_2(D_{p,q})$ with entrees in the ring $D_{p,d}$ and a translationally invariant phase
valued function $\varphi\mathpunct: D_{p,d}^2\to \8U(1)$ such that 
\begin{equation}
\label{eq:CQCA}
\alpha(\1w(\xi))=\varphi(\xi)\1w(\1s\star\xi)
\end{equation}
and $\varphi$ fulfills the cocycle condition
\begin{equation}
\label{eq:COCYCLE}
\varphi(\xi+\eta)=\epsilon_p^{\beta(\xi,\eta)-\beta(\1s\star\xi,\1s\star\eta)}
\varphi(\xi)\varphi(\eta) \; .
\end{equation}
Moreover, the map $\1s\star$ preserves the symplectic form $\sigma$,
i.e. $\sigma(\1s\star\xi,\1s\star\eta)=\sigma(\xi,\eta)$. 
\end{theorem}

We sketch here just the basic idea of the proof: It follows from the Weyl relations that each
automorphism $\alpha$ that maps Weyl operator to multiples of Weyl operators according to
(\ref{eq:CLIFF}) induces a $\7F_p$-linear map $\1S$ on phase space that preserves the symplectic
form. Moreover, the condition to be an automorphism implies that the phase valued function
$\varphi$ fulfills (\ref{eq:COCYCLE}) where, for this moment, we have to repace the operator
$\1s\star$ by $\1S$. 

By taking advantage of the translation invariance, the phase valued function $\varphi$ is
translationally invariant and the symplectic map $\1S$ commutes with the lattice translations, it
follows that $\1S$ is given by the convolution $\1s\star$ with a matrix-valued function with finite
support. Note that a two-by-two matrix $\1s$ with entries in the ring $D_{p,d}$ can equivalently be
seen as a function that maps a lattice site $x\in\7Z^d$ to a two-by-two matrix with entries in the
finite field $\7F_p$. The convolution with a phase space $\xi$ vector is given by
$(\1s\star\xi)(x)=\sum_y\1s(y)\cdot\xi(x-y)$ where $\cdot$ is usual matrix multiplication. The
localization region of a Weyl operator $\1w(\xi)$ is just the support of the function $\xi$. If the
support of $\xi$ is just a single site $x$, then $\1s\star \xi$ has support in $N+x$, where $N$ is
the support of $\1s$. Therefore, the application of the corresponding CQCA yields an operator
$\alpha(\1w(\xi))=\varphi(\xi)\1w(\1s\star\xi)$ which is localized in $N+x$. As a consequence, the
support
of $\1s$ determines the neighborhood scheme of the QCA.

We also shown in \cite{SchlVogtWer08} that each matrix-valued function $\1s$ for which the
convolution $\1s\star$ preserves the symplectic form $\sigma$ a phase valued function $\varphi$ can
be found, such that the equation (\ref{eq:CQCA}) defines a CQCA. Therefore, the classification of
CQCAs is equivalent to characterize all two-by-two matrices $\1s\in\8M_2(D_{p,d})$ with entries in
the ring $D_{p,d}$ whose convolution $\1s\star$ is symplectic. In accordance with
\cite{SchlVogtWer08}, we call the convolution $\1s\star$ a ``symplectic cellular automaton (SCA)''.

\section{On the structure of Clifford quantum cellular automata}
For the further analysis of CQCAs, we have a closer look at the ring and module structure of the
phase space $\Xi_{p,d}=D_{p,d}^2$. As already mentioned $\Xi_{p,d}$ is a two dimensional module
over the ring $D_{p,d}$ where the product is the convolution. A function $f\in D_{p,d}$ acts on a
phase space vector by $f\star\xi=f\star(\xi_+,\xi_-)=(f\star\xi_+,f\star\xi_-)$. A symplectic
cellular automaton (SCA), which induces a CQCA, is then a module homomorphism. Namely, since the
convolution in $D_{p,d}$ is commutative, we observe that $\1s\star f\star \xi= f\star\1s\star\xi$. 

To take advantage of the translation symmetry in an appropriate manner, we have introduced the
``algebraic Fourier transform'', which identifies the ring $D_{p,d}$ with the commutative ring of
Laurent polynomials 
\begin{equation}
\hat D_{p,d}=\7F_p[u_1,u_2,\cdots,u_d,u_1^{-1},\cdots, u_d^{-1}]
\end{equation}
generated by the variables $u_1,\cdots,u_d$ and its inverses $u_1^{-1},\cdots,u_d^{-1}$. For a
function $f\in D_{p,d}$ the corresponding Laurent-polynomial is simply given by
\begin{equation}
\hat f=\sum_x f(x) u^x
\end{equation}
where we write $u^x=u_1^{x_1}u_2^{x_2}\cdots u_d^{x_d}$ for a handy notation. To view the elements
in $D_{p,d}$ as formalpolynomials, gives us a convenient book-keeping at hand. Namely, the
convolution turns into a product
of polynomials, i.e. for two functions the identity  
\begin{equation}
\widehat{f\star h}=\hat f \hat h
\end{equation}
holds. We mention here that the ring $\hat D_{p,d}$ is a ``divison ring'' which means that $fh=0$
implies that either $f=0$ or $h=0$. Morover, the only invertible elements in $\hat D_{p,d}$ are the
monomials $u^x$ with $x\in\7Z^d$. 

With help of this ring isomorphism $f\mapsto \hat f$, the phase space can be identified with $\hat
D_{p,d}^2$ and, since a SCA is a module homomorphism, its Fourier transform just acts by matrix
multiplication. To be more precise, a SQCA is given by a two-by-two matrix $\hat{\1s}$ with entries
in the polynom ring $\hat D_{p,d}$ acting on a phase space vector by
\begin{equation}
\widehat{\1s\star\xi}
=\left(\begin{array}{cc}\hat{\1s}_{++}&\hat{\1s}_{+-}\\
                              \hat{\1s}_{-+}&\hat{\1s}_{--}
                             \end{array}\right)\left(\begin{array}{c}\hat\xi_+\\
                              \hat\xi_-
                              \end{array}\right)
=
\left(\begin{array}{c}\hat{\1s}_{++}\hat\xi_++\hat{\1s}_{+-}\hat\xi_-\\
                              \hat{\1s}_{-+}\hat\xi_++\hat{\1s}_{--}\hat\xi_-
                              \end{array}\right) \; .
\end{equation}

After applying the algebraic Fourier transform, the symplectic form is a $\hat D_{p,d}$-valued
bilinear from $\Sigma$ on $\hat D^2_{p,d}$ that can be calculated according to 
\begin{equation}
\Sigma(\xi,\eta)=\overline{\xi_+}\eta_--\overline{\xi_-}\eta_+ \; ,
\end{equation}
where $f\mapsto\overline{f}$ is the involution on the ring $\hat D_{p,d}$ which replaces
in the polynomial $f=f(u)$ the variable $u_k$ by its inverse $u_k^{-1}$. In the lattice space, this
corresponds to a reflection at the origin. The form $\Sigma$ is related to the underlying
symplectic form $\sigma$ by 
\begin{equation}
\Sigma(\xi,\eta)=\sum_x\sigma(\check\xi,\tau_x\check\eta)u^x 
\end{equation}
where $f\mapsto \check f$ is the inverse algebraic Fourier transform sending a polynomial $f$ to a
function $\check f$ on the lattice. It is not difficult to observe, that $\Sigma$ is a module
homomorphism in the second argument, i.e. $\Sigma(\xi,f\eta)=\Sigma(\xi,\eta)f$ and
that it fulfills the relation
$\Sigma(\xi,\eta)=-\overline{\Sigma(\eta,\xi)}=-\Sigma(\overline{\eta},\overline{\xi})$. Thus
$\Sigma$ is antisymmetric for reflection invariant polynomials. In this context, a helpful lemma
for the characterization of SCAs is the following: 

\begin{theorem}{Lemma}
A two-by-two matrix $\1s\in \8M_2(\hat D_{p,d})$ with entries in the polynom ring $\hat
D_{p,d}$ is a symplectic cellular automaton, if and only if, it preserves the form $\Sigma$,
i.e. the identity $\Sigma(\1s\xi,\1s\eta)=\Sigma(\xi,\eta)$ holds.
\end{theorem}

According to this lemma, the characterization of CQCAs (hence SCAs) reduces to the problem of
finding two-by-two matrices with entries in the ring $\hat D_{p,d}$ which preserve $\Sigma$.
Since we deal here with module homomorphism --- a linear structure over the ring
$\hat D_{p,d}$ --- we have reduced a problem in infinitely many degrees of freedom to an
effectively two-dimensional problem.  

There is an important subring in $\hat D_{p,d}$, which we denote here by $P_{p,d}$,which consists
of all polynomials that are invariant under the reflection $u_k\mapsto u_k^{-1}$ which means
$f=f(u)=f(u^{-1})=\overline{f}$. For $d=1$, the corresponding function $\check f$ in lattice space
is then given by a palindrome string $\check f=(q_n q_{n-1} \cdots q_1 q_0 q_1q_2\cdots q_n)$
starting at the left boundary of the support $x=-n$ and ending at the right boundary $x=n$. For this
reason, we call $P_{p,d}$ the polynom subring of palindromes in $\hat D_{p,d}$. If we look at the
properties of the form, $\Sigma$, we see that it is a non-degenerate antisymmetric
$P_{p,d}$-bilinear form on $P_{p,d}^2$. In analogy, that the symplectic group of $\7C^2$ is given by
the special linear group $\8{SL}(2,\7C)$ a first guess is, that the group of SCA is given by all
two-by-two matrices with entries in the palindrome subring $P_{p,d}$ having ring-valued determinant
equal to one. Indeed, if we choose a matrix $\1s\in\8{SL}(2,P_{p,d})$, then we observe by a
straight forward calculation that $\1s$ preserves the form  $\Sigma$. If we multiply $\1s$ with a
monomial $u^a$, which corresponds to a lattice translation by $a\in\7Z^d$, then we observe that
$\Sigma(u^a\1s\xi, u^a\1s\eta)=\Sigma(\1s\xi,\1s\eta)u^{-a}u^a=\Sigma(\xi,\eta)$. Thus if $\1s$ is
a SCA then $u^a\1s$ is a SCA too. Indeed, all SCAs are of this type. The precise statement which we
have established in \cite[Theorem~3.4]{SchlVogtWer08}
is the following:

\begin{theorem}{Theorem}
\label{thm:main}
The group of Clifford quantum cellular automata acting on a $d$-dimensional lattice with single
cell algebras $\8M_p(\7C)$ is isomorphic to the direct product $\7Z^d\times\8{SL}(2,P_{p,d})$ of
the lattice translation group and the special linear group of two-by-two matrices with entries in
in the palindrome subring $P_{p,d}$. 
\end{theorem}

This theorem can be used to build up a simple cooking recipe for constructing CQCAs. Firstly,
take two palindromes $f,h\in P_{p,d}$. Recall that palindromes are easy to get. Namely, for the case
that $g$ is not a palindrome you just make one by taking $h=g+\overline{g}$. Secondly, factorize the
polynomial $1-fg=f'h'$ in the subring $P_{p,d}$ into two palindromes $f',h'$. Finally, you get
your CQCA by building the matrix
\begin{equation}
\1s=\left(\begin{array}{cc}
           f&f'\\ h'&h
          \end{array}\right) \; .
\end{equation}
For this type of recipe, to find all possible factorizations of the polynomial $1-fh$ is the
crucial problem which can be quite cumbersome. However, there is allways the trivial solution which
is given by $h'=1$ and $f'=1-fh$. 

Concerning the factorization problem, at least for a one dimensional lattice $d=1$ the situation can
be tackled. Here one takes advantage of the fact that the ring $P_{1,p}$ is a so called ``Euclidean
ring'' (see e.g. \cite{Jac75}) and an extended Euclidean algorithm for finding greatest common
divisors can be applied. This yields in an even stronger classification result of one-dimensional
CQCAs than provided by the dimension independent Theorem~\ref{thm:main}. We have shown the
following 
\cite[Theorem 3.11]{SchlVogtWer08}:

\begin{theorem}{Theorem}
\label{thm:one-dim}
Every Clifford quantum cellular automata $\1s$ acting on a one dimensional lattice with single
cell algebras $\8M_p(\7C)$ can be factorized into a product of a unique shift $u^a$ and elementary
CQCAs of the following two types: The first type is a shear transformation 
\begin{equation}
\1g_n=\left(\begin{array}{cc}
           1&0\\ u^{-n}+u^n&1
          \end{array}\right)
\end{equation}
depending on an integer $n\in\7N$. The second type depends on a constant $c\in \7F_p$ according to 
\begin{equation}
\1f_c=\left(\begin{array}{cc}
           0&c\\ -c^{-1}&0
          \end{array}\right) \; .
\end{equation}
\end{theorem}

The automata $\1f_c$ are constant matrix valued polynomials which implies that they act on each
lattice site independently. The have zero propagation speed since their neighborhood scheme only
consits of the origin. On the other hand the propagation of the localization region of an
observable is induced by the shear automata $g_n$. Their local rule propagate from the origin into
the cells $\{-n,n\}$.

\section{Concluding remarks}
In this note, we have discussed some aspects on the structure of quantum cellular automata where we
have mainly focused on our results on Clifford quantum cellular automata \cite{SchlVogtWer08}.

We have characterized the group of CQCAs in terms of symplectic cellular automata on a suitable
phase space. With the help of the concept of algebraic Fourier transform, this phase space can be
identified with two-dimensional vectors of Laurent-polynomials, and symplectic cellular automata can
be described by two-by-two matrices with Laurent-polynomial entries. We have reported that
these entries must be reflection invariant and that up to some global shift the
determinant of the matrix must be one, so the group of CQCAs is isomorphic to the direct
product of the lattice translation group with the special linear group of two-by-two matrices with
reflection invariant polynomials as matrix elements.

Due to the specialty that for a 1D lattice we are faced with an Euclidean ring,  each
one-dimensional CQCA can be factorized into a product of elementary shear automata and local
transforms.

Besides the core results, that we have presented here, there is a correspondence between 1D CQCAs
and 1D translationally invariant stabilizer (graph) states (see e.g. \cite{Got97,Schl03} for the
notion of stabilizer (graph) states). For a fixed translationally invariant pure stabilizer state,
which is in particular a product state, every other translationally invariant pure stabilizer state
can be created by applying an appropriate CQCA.

A further natural question is concerned with lattices with periodic boundary conditions. Here the
techniques from infinitely extended lattices can be applied to a certain extend. The technical
problem is here, however, that the involved polynom ring is no longer a division ring.

\newpage



\begin{thebibliography}{10}

\bibitem{BerVaz93}
E.~Bernstein and U.~Vazirani.
\newblock Quantum complexity theory.
\newblock In {\em Proc. 25th Ann. ACM Symp. on Theory of Computing}, pages
  11--20, 1993.

\bibitem{Conw76}
J.~H. Conway.
\newblock {\em On numbers and games}.
\newblock Lect. Notes Pure Appl. Math. Academic Press, London, 1976.

\bibitem{Fey82}
R.~Feynman.
\newblock Simulating physics with computers.
\newblock {\em Int. J. Theor. Phys.}, 21:467--488, 1982.

\bibitem{Got97}
D.~Gottesman.
\newblock Stabilizer codes and quantum error correction.
\newblock Ph.D. thesis, Caltech, quant-ph/9705052, 1997.

\bibitem{Hol04}
A.~S. Holevo.
\newblock Additivity conjecture and covariant channels.
\newblock In {\em Proc. Conference "Foundations of Quantum Information"},
  Camerino, 2004.

\bibitem{Jac75}
N.~Jacobson.
\newblock {\em Lectures in abstract algebra. Volume I: Basic Concepts}, volume
  Graduate Texts in Mathematics, No. 30.
\newblock Springer-Verlag, New-York, Berlin, 1975.

\bibitem{Schl03}
D.-M. Schlingemann.
\newblock Cluster states, graphs and algorithms.
\newblock {\em Quant. Inf. Comp.}, 4:287--324, 2004.

\bibitem{SchlVogtWer08}
D.-M. Schlingemann, H.~Vogts, and R.~F. Werner.
\newblock On the structure of clifford quantum cellular automata.
\newblock {\em J. Math. Phys.}, 49:112104, 2008.

\bibitem{SchuWer04}
B.~Schumacher and R.~F. Werner.
\newblock Reversible quantum cellular automata.
\newblock quant-ph/0405175, 2004.

\bibitem{Scu79}
H.~Scutaru.
\newblock Some remarks on covariant completely positive maps.
\newblock {\em Rep. Math. Phys.}, 16:79--87, 1979.

\bibitem{Wat95}
J.~Watrous.
\newblock On one-dimensional quantum cellular automata.
\newblock In {\em Proc. 36th Ann. Symp. on Foundations of Computer Science},
  pages 528--537, 1995.

\bibitem{Zmud72}
E.~M. Zmud.
\newblock Symplectic geometry and projective representations of finite abelian
  groups.
\newblock {\em Math. USSR Sbornik}, 16:1--16, 1972.

\end{thebibliography}
\end{document}